\documentclass[aps,pra,twocolumn,amsmath,amssymb,nofootinbib,superscriptaddress]{revtex4}

\newcommand{\ket}[1]{|#1\rangle}

\usepackage[pdftex]{graphicx}
\usepackage{mathrsfs}
\usepackage{csquotes}
\usepackage[colorlinks]{hyperref}

\begin{document}

\bibliographystyle{apsrev}

\title{Boson sampling with displaced single-photon Fock states versus single-photon-added coherent states\textemdash\\The quantum-classical divide and computational-complexity transitions in linear optics}

\author{Kaushik P. Seshadreesan}
\email{ksesha1@lsu.edu}
\affiliation{Hearne Institute for Theoretical Physics and Department of Physics \& Astronomy, Louisiana State University, Baton Rouge, LA 70803}

\author{Jonathan P. Olson}
\email{jolson7@lsu.edu}
\affiliation{Hearne Institute for Theoretical Physics and Department of Physics \& Astronomy, Louisiana State University, Baton Rouge, LA 70803}

\author{Keith R. Motes}
\affiliation{Centre for Engineered Quantum Systems, Department of Physics and Astronomy, Macquarie University, Sydney NSW 2113, Australia}

\author{Peter P. Rohde}
\email[]{dr.rohde@gmail.com}
\homepage{http://www.peterrohde.org}
\affiliation{Centre for Engineered Quantum Systems, Department of Physics and Astronomy, Macquarie University, Sydney NSW 2113, Australia}
\affiliation{Centre for Quantum Computation and Intelligent Systems (QCIS), Faculty of Engineering \& Information Technology, University of Technology, Sydney, NSW 2007, Australia}

\author{Jonathan P. Dowling}
\affiliation{Hearne Institute for Theoretical Physics and Department of Physics \& Astronomy, Louisiana State University, Baton Rouge, LA 70803}
\affiliation{Computational Science Research Center, Beijing 100084, China}

\date{\today}

\frenchspacing

\begin{abstract}

Boson sampling is a specific quantum computation, which is likely hard to implement efficiently on a classical computer. The task is to sample the output photon number distribution of a linear optical interferometric network, which is fed with single-photon Fock state inputs. A question that has been asked is if the sampling problems associated with any other input quantum states of light (other than the Fock states) to a linear optical network and suitable output detection strategies are also of similar computational complexity as boson sampling. We consider the states that differ from the Fock states by a displacement operation, namely the displaced Fock states and the photon-added coherent states. It is easy to show that the sampling problem associated with displaced single-photon Fock states and a displaced photon number detection scheme is in the same complexity class as boson sampling for all values of displacement. On the other hand, we show that the sampling problem associated with single-photon-added coherent states and the same displaced photon number detection scheme demonstrates a computational complexity transition. It transitions from being just as hard as boson sampling when the input coherent amplitudes are sufficiently small, to a classically simulatable problem in the limit of large coherent amplitudes.%The photon-added coherent states could thus be potentially used to study the classical simulatability/non-simulatability boundary similar to the quantum/classical boundary.

%The primary idea of boson sampling is that sampling the output distribution of a linear optical multimode interferometer fed with single photon Fock states is a problem likely hard to simulate using a classical computer.   We also show that the photon-added coherent state s are computational complexity when evolved using linear optical networks and measured using {\it on-off} photodetection. We conWe show that provided that the coherent state amplitudes are upper bounded by an inverse polynomial in the size of the system, the sampling problem remains computationally hard.
\end{abstract}

\maketitle

\section{Introduction}
Quantum computation promises the ability to solve problems intractable on classical computers, including fast integer factorization~\cite{Shor_97}, quantum search algorithms~\cite{Grover_96}, and quantum simulation applications~\cite{Lloyd_96}. There exist many models of universal quantum computation~\cite{bib:NielsenChuang00}. The Knill, Laflamme and Milburn (KLM) proposal for linear optical quantum computation (LOQC)~\cite{bib:KLM01} is a well known example of such a model. However, the technological challenges to realize scalable, full-fledged LOQC (or any other existing model of quantum computation for that matter) continue to remain daunting.%In doing so, it challenges the extended Church-Turing thesis (ECT) of the theory of computation~\cite{Aaronsonbook}, which is the claim that any computation that is performed on a physical device can be efficiently simulated on a classical computer with only a polynomial overhead. There are many potential candidates for implementing scalable quantum computation. Linear optical quantum computation (LOQC) \cite{bib:KLM01} is one of the leading candidates owing to the long decoherence times of photons, and the relative ease to prepare, interfere and detect them. However, the technological requirements for full-fledged LOQC are daunting, stimulating interest in developing simpler approaches for LOQC.

In 2011, Aaronson \& Arkhipov (AA) introduced and studied a specific computational task based on linear optics, called boson sampling~\cite{bib:AaronsonArkhipov10}. The task of boson sampling is to sample the output photon number distribution of a linear optical network, which is fed with single photons. AA showed that boson sampling is likely hard to implement efficiently on a classical computer, but can be efficiently implemented on a quantum device. Also, boson sampling is less demanding to realize experimentally than LOQC. As a result, boson sampling has garnered much interest. Many experimental demonstrations of boson sampling have been carried out~\cite{bib:Broome2012, bib:Spring2, bib:Tillmann4, bib:Crespi3}. Special efforts are underway towards scalable implementations of boson sampling~\cite{bib:Motes13, Lund_13, MGDR14, Shen_14}. Ways to certify true boson sampling in order to distinguish it from uniform sampling, classical sampling, or random-state sampling have been developed~\cite{bib:AA13response, Spagnolo_13, Carolan_13, bib:Molmer13}. The effects of realistic implementations of boson sampling such as mode-mismatch, spectra of the bosons, and spectral sensitivities of detectors, have been studied~\cite{Shc14a, Rho14}. This has lead to a theory of interference with partially indistinguishable particles~\cite{Shc14b, Tich14}. Recently, for the first time, boson sampling (with a suitably modified input state) has been shown to yield a practical tool for difficult molecular computations to generate molecular vibronic spectra~\cite{HGPMA14}. 

Another question that has been investigated is whether there are quantum states of light---other than the Fock states---which when evolved through a linear-optical circuit and sampled using a suitable detection strategy, also implement likely classically hard problems similar to AA's boson sampling. Recent results have shown that, in the case of Gaussian states (most generally displaced, squeezed, thermal states), sampling in the photon number basis can be just as hard as boson sampling~\cite{bib:Lund13}. To further elaborate, while the sampling of thermal states can be simulated efficiently by a classical algorithm~\cite{RLR_14}, it has been shown that the sampling of squeezed vacuum states is likely hard to efficiently simulate classically at least in some special cases~\cite{bib:PhysRevA.88.044301, bib:Lund13}. Among non-Gaussian inputs (other than the Fock states), the photon-added and subtracted squeezed vacuum states~\cite{bib:Olson14} and generalized cat states (arbitrary superpositions of coherent states)~\cite{bib:RohdeCat}, along with photon number detection, have been shown to likely implement computationally hard sampling problems similar to boson sampling.

In this article, we study the linear optics-based sampling problems associated with the quantum states of light that differ from the Fock states by the displacement operation of quantum optics, namely the displaced Fock states and the photon-added coherent states (which are the photon-added displaced vacuum states), and a displaced photon number detection. The displacement operator can be written as
\begin{equation}
\label{disop}
\hat{D}(\alpha)=\exp\left(\alpha \hat{a}^{\dagger}-\alpha^*\hat{a}\right),
\end{equation}
where $\alpha$ is a complex amplitude that quantifies displacement in phase space, and $\hat{a}^\dagger$ is the mode photon creation operator. The displaced single-photon Fock state (DSPFS) is the state $\hat{D}(\alpha)\hat{a}^\dagger|0\rangle$, while the single-photon-added coherent state (SPACS) is $\propto\hat{a}^\dagger\hat{D}(\alpha)|0\rangle$. Although these input states are in practice more difficult to prepare than the single-photon Fock state, the associated sampling problems allow us to demonstrate a transition in the computational complexity of linear optics. It is easy to show that the DSPFS sampling problem is in the same complexity class as AA boson sampling for any displacement $\alpha$. However, the SPACS, differing only in the ordering of the operators, presents an interesting case---we show that the sampling problem with SPACS is just as hard as AA's boson sampling when the input coherent amplitudes are sufficiently small (subject to a bound that we derive explicitly), but transitions into a problem that is easy to simulate classically in the limit of large input coherent amplitudes.

In Sec.~\ref{originalbs}, we briefly review AA's original scheme for boson sampling. In Secs.~\ref{dfssampling} and \ref{pacssampling}, we discuss the sampling problems associated with the DSPFS and the SPACS, respectively. In Sec.~\ref{concl}, we state some conclusions and summarize our work.%In Sec.~\ref{disc}, we discuss about how the sampling of single-photon PACS could be used to explore the classical simulatability/non-simulatability boundary, and conclude with a brief summary of our findings.

\section{Aaronson and Arkhipov's boson sampling}
\label{originalbs}
In boson sampling as presented by AA~\cite{bib:AaronsonArkhipov10}, single-photon Fock states $\ket{1}$ are fed into the first $n$ modes of an $m$-mode linear-optical interferometer where $m=\Omega(n^2)$. The remaining \mbox{$m-n$} modes are injected with the vacuum state $\ket{0}$. The overall input state is therefore
\begin{eqnarray} \label{eq:input_state}
\ket{\psi_\mathrm{in}} &=& \ket{1_1,\dots,1_n,0_{n+1},\dots,0_m} \nonumber \\
&=& \left(\prod_{i=1}^{n}\hat{a}_i^\dagger\right) \ket{0_1,\dots,0_m},
\end{eqnarray}
where $\hat{a}_i^\dag$ is the photon creation operator of the $i$th mode. The interferometer, which is comprised of a network of $O(m^2)$ number of beamsplitters and phase-shifters, then implements an \mbox{$m\times m$} unitary operation $\hat{U}$ on the input state~\cite{bib:Reck94}. %Passive unitary operators are those unitaries which preserve the total number of photons from input to output. (It was shown by Reck \emph{et al.} \cite{bib:Reck94} that any \mbox{$m\times m$} passive unitary map may be efficiently implemented using $O(m^2)$ optical elements.) 
Under the action of the unitary, the photon-creation operators of the modes transform as 
\begin{equation} \label{eq:unitary_map}
\hat{U}\hat{a}_i^\dag\hat{U}^\dagger = \sum_{j=1}^m U_{i,j} \hat{a}_j^\dag.
\end{equation}
This results in an output state, which is a superposition of every possible $n$-photon-number configuration
\begin{equation} \label{eq:output_dist}
\ket{\psi_\mathrm{out}}^{\mathrm{AA}} = \sum_S \gamma_S \ket{s_1^{(S)},\dots,s_m^{(S)}},
\end{equation}
where $S$ represents an output configuration of the $n$ photons, $\gamma_S$ is the amplitude associated with configuration $S$, and $s_i^{(S)}$ is the number of photons in mode $i$ at the output associated with configuration $S$. (Note that we suppress the superscript $(S)$ hereafter for brevity of notation.) The number of configurations is
\begin{equation} \label{eq:terms}
|S| = \binom{n+m-1}{n},
\end{equation}
which grows exponentially with $n$. Finally, the probability distribution $P(S)=|\gamma_S|^2$, is sampled using coincidence photon number detection (CPND). For a large $n$, AA conjecture that the number of modes $m=\Omega(n^2)$ sufficiently ensures that, to a high probability, no more than a single photon arrives per mode at the output. (This is sometimes referred to as the \enquote{bosonic birthday paradox}.) Therefore, on-off photodetectors are sufficient to perform the measurements for large instances of the problem. The protocol is post-selected upon measuring all $n$ photons at the output; \emph{i.e.} so that the total photon number is preserved $\sum_i s_i^{(S)} = n\,\,\forall\, S$. This accounts for any potential losses due to inefficiencies in the experimental devices. 

The statistics of the output distribution $P(S)$ are sampled by repeated application of the protocol. The output amplitudes are given by $\gamma_S \propto \mathrm{Per}(U_{S,T})$, where \mbox{$U_{S,T}$} is a \mbox{$n\times n$} submatrix of $U$ given as a function of the fixed input configuration $T=\left(1_1,\dots,1_n,0_{n+1},\dots,0_m\right)$ and the output configuration $S$. Since calculating the permanent of an arbitrary complex-valued matrix is \textbf{\#P}-complete, requiring $O(2^n n^2)$ runtime according to Ryser's algorithm \cite{bib:Ryser63}, boson sampling is believed to be likely hard to simulate classically. Also, exact boson sampling by a polynomial-time classical probabilistic algorithm would imply a collapse of the polynomial hierarchy, while non-collapse of the polynomial hierarchy is generally believed to be a reasonable conjecture~\cite{bib:AaronsonArkhipov10}. Gard \emph{et al.} gave an elementary argument that classical computers likely cannot efficiently simulate multimode linear-optical interferometers with arbitrary Fock-state inputs \cite{bib:gard2013classical}. AA~\cite{bib:AaronsonArkhipov10} gave a full complexity proof for the exact case where the device is noiseless. For the noisy case, a partial proof was provided which requires two conjectures that are believed likely to be true. A detailed elementary introduction to boson sampling was presented in~\cite{GMORD_14}. Our results are based on the assumption that both exact and approximate boson-sampling are classically hard problems, which is highly likely the case following the work of AA.

%Although there are no known applications for boson sampling, the mere fact that this sampling problem is classically intractable makes it of great interest. Note that boson sampling is \emph{not} an algorithm for \emph{calculating} given matrix permanents, as this would require an exponential number of samples. We will use AA's result for approximate boson sampling later when deriving our bounds. 

\section{Sampling Displaced Fock states}
\label{dfssampling}

Consider the DSPFS in place of the single-photon Fock states in (\ref{eq:input_state}) as inputs to a linear-optical interferometer. That is, consider an overall input state of the form
\begin{equation} \label{eq:input_state_DFS}
\ket{\psi_\mathrm{in}}^{\mathrm{DSPFS}} =\left( \prod_{i=1}^{n}\hat{D}_i\left(\alpha^{(i)}\right)\hat{a}_i^\dagger\right) \ket{0_1,\dots,0_m},
\end{equation}
where $\hat{D}_i\left(\alpha^{(i)}\right)$ is the displacement operator of the $i$th mode, and $\alpha^{(i)}$ is the complex coherent amplitude for the displacement. A unitary operation $\hat{U}$ then transforms the state into $|\psi_{\rm out}\rangle^{\mathrm{DSPFS}}$
\begin{align}
\label{UxformDFS}
%&=\hat{U}|\psi_{\rm in}\rangle^{\mathrm{DFS}}\nonumber\\
&=\hat{U} \left(\prod_{i=1}^{n}\hat{D}_i\left(\alpha^{(i)}\right)\hat{a}_i^\dagger\right)\hat{U}^{\dagger}\hat{U}\ket{0_1,\dots,0_m},\nonumber\\
&=\hat{U} \left(\prod_{i=1}^{n}\hat{D}_i\left(\alpha^{(i)}\right)\right)\hat{U}^{\dagger}\hat{U}\left(\prod_{k=1}^{n}\hat{a}_k^\dagger\right)\hat{U}^{\dagger}\ket{0_1,\dots,0_m}\nonumber\\
&= \prod_{i=1}^{n}\left(\hat{U}\hat{D}_i\left(\alpha^{(i)}\right)\hat{U}^{\dagger}\right)\prod_{k=1}^{n}\left(\hat{U}\hat{a}_k^\dagger\hat{U}^{\dagger}\right)\ket{0_1,\dots,0_m}\nonumber\\
&=\left(\prod_{j=1}^{m}\hat{D}_j\left(\beta^{(j)}\right)\right)\left(\sum_{S}\gamma_S (\hat{b}_1^{\dagger})^{s_1}(\hat{b}_2^{\dagger})^{s_2}\dots(\hat{b}_m^{\dagger})^{s_m}\right)\nonumber\\
&\times\ket{0_1,\dots,0_m},
\end{align}
where $\beta^{(j)} = \sum_i U_{i,j} \alpha^{(i)}$ is the new displacement amplitude in the $j$th mode, $\hat{b}_k^{\dagger}$ is the photon-creation operator of the  $k$th mode, and $s_k$ is the number of photons in the $k$th mode, associated with configuration $S$ at the output such that $\sum_{k=1}^m s_k=n$ for each $S$. In deriving (\ref{UxformDFS}), we have used the following: $\hat{U}^{\dagger}\hat{U}=I$, $\hat{U}\ket{0_1,\dots,0_m}=\ket{0_1,\dots,0_m}$, (\ref{eq:unitary_map}) and (\ref{eq:output_dist}), and the fact that the action of a unitary on a tensor product of coherent states results in another tensor product of coherent states as shown in Appendix A of~\cite{bib:RohdeCat}. The final expression is nothing but a displaced version of the output state of AA's boson sampling given in (\ref{eq:output_dist}).

For any unitary operator $U$, the new complex displacement amplitudes $\beta^{(j)}$ can be efficiently computed. Since $D(-\alpha)D(\alpha)=I$, a counter-displacement with amplitudes $-\beta^{(j)}$ could be applied to the $m$ output modes. The displacement operation could be performed using unbalanced homodyning~\cite{BW99,WLD07}. Upon such a displacement operation, the sampling problem associated with the output state reduces to AA's boson sampling given in (\ref{eq:output_dist}), which 
%. We once again assume that the interferometer is large, with the number of modes being $m=O(n^2)$, so that $s_k=\{0,1\}$ in (\ref{UxformDFS}). This
can subsequently be accessed using CPND. Thus, the linear-optics sampling of the DSPFS with our modified measurement scheme at the output comprising of an inverse displacement followed by CPND is in the same complexity class as AA's boson sampling. While this observation may appear trivial---since a product of displacement operators commutes through a linear-optical network to yield another product of displacement operators---it demonstrates that an entire class of quantum states of light yield a problem of equal complexity to boson sampling, with a suitable adaptation of the measurement scheme.%This is not surprising, since the displacement operator, akin to the boson sampling unitary, is also a passive unitary transformation, and therefore doesn't change the computational complexity of the sampling problem.

\section{Sampling photon-added coherent states}
\label{pacssampling}

Now consider the SPACS instead of the DSPFS. These states differ from the DSPFS only in the ordering of the operators. However, since the displacement operator of (\ref{disop}) does not commute with the photon creation operator $\hat{a}^\dagger$, the SPACS and the DSPFS are distinctly different states. 
%We will now show that the ordering of the operators brings about a difference in the computational complexity of the two systems. 

A $k$-photon-added coherent state may be written as
\begin{equation}
\label{pacsdefinition}
|\alpha,k\rangle=\mathcal{N}_k \hat{a}^{\dagger^{k}}|\alpha\rangle,
\end{equation}
where $\alpha$ is the complex coherent amplitude and the normalization is
\begin{equation}
\mathcal{N}_k=\frac{1}{\sqrt{k!L_k (-|\alpha|^2)}},
\end{equation}
$L_k$ being the Laguerre polynomial of order $k$. Such states were first discussed by Agarwal \& Tara \cite{Agarwal_91}. The SPACS we consider here thus corresponds to $\ket{\alpha,1}$ of (\ref{pacsdefinition}). 

\begin{figure}
\includegraphics[width=0.5\columnwidth]{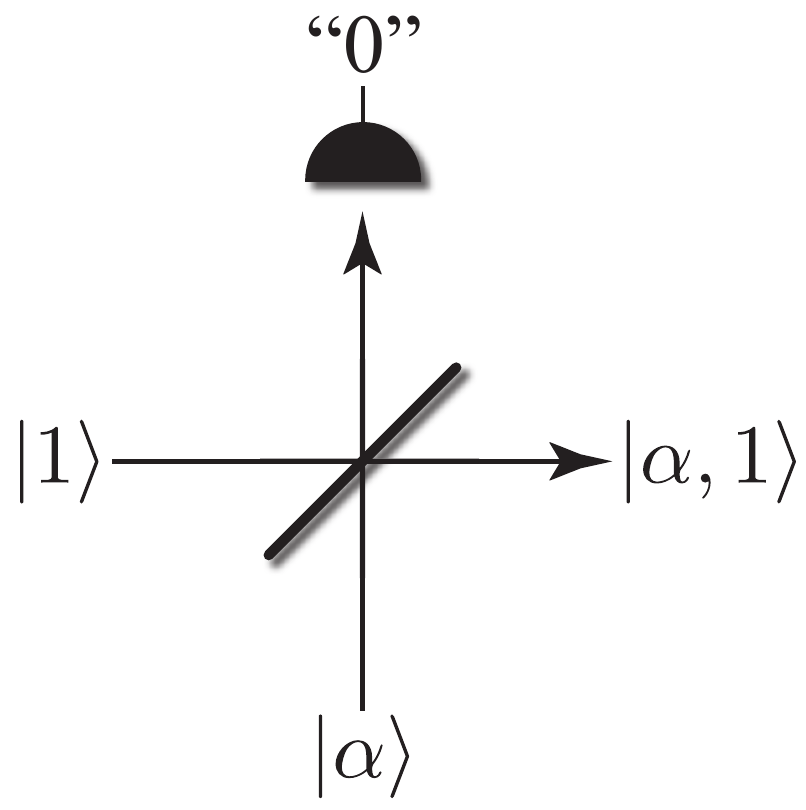}
\caption{ When a coherent state and a single photon state are mixed on a highly reflective beamsplitter, and no photon is detected in the transmitted mode, a SPACS is heralded in the transmitted mode.} \label{fig:PACS_prep}
\end{figure}

Consider a scheme where a single photon (e.g. prepared via heralded spontaneous parametric down-conversion) is mixed with a coherent state on a highly reflective beam splitter (Fig.~\ref{fig:PACS_prep}). When a single-photon detector placed in the transmitted mode detects vacuum, we know that the incident photon has been emitted into the other output port, and thus a SPACS has been heralded~\cite{Dakna_98, Dakna_98_2, Zavatta_04, Zavatta_05}.

\begin{figure}[!htb]
\includegraphics[width=0.5\columnwidth]{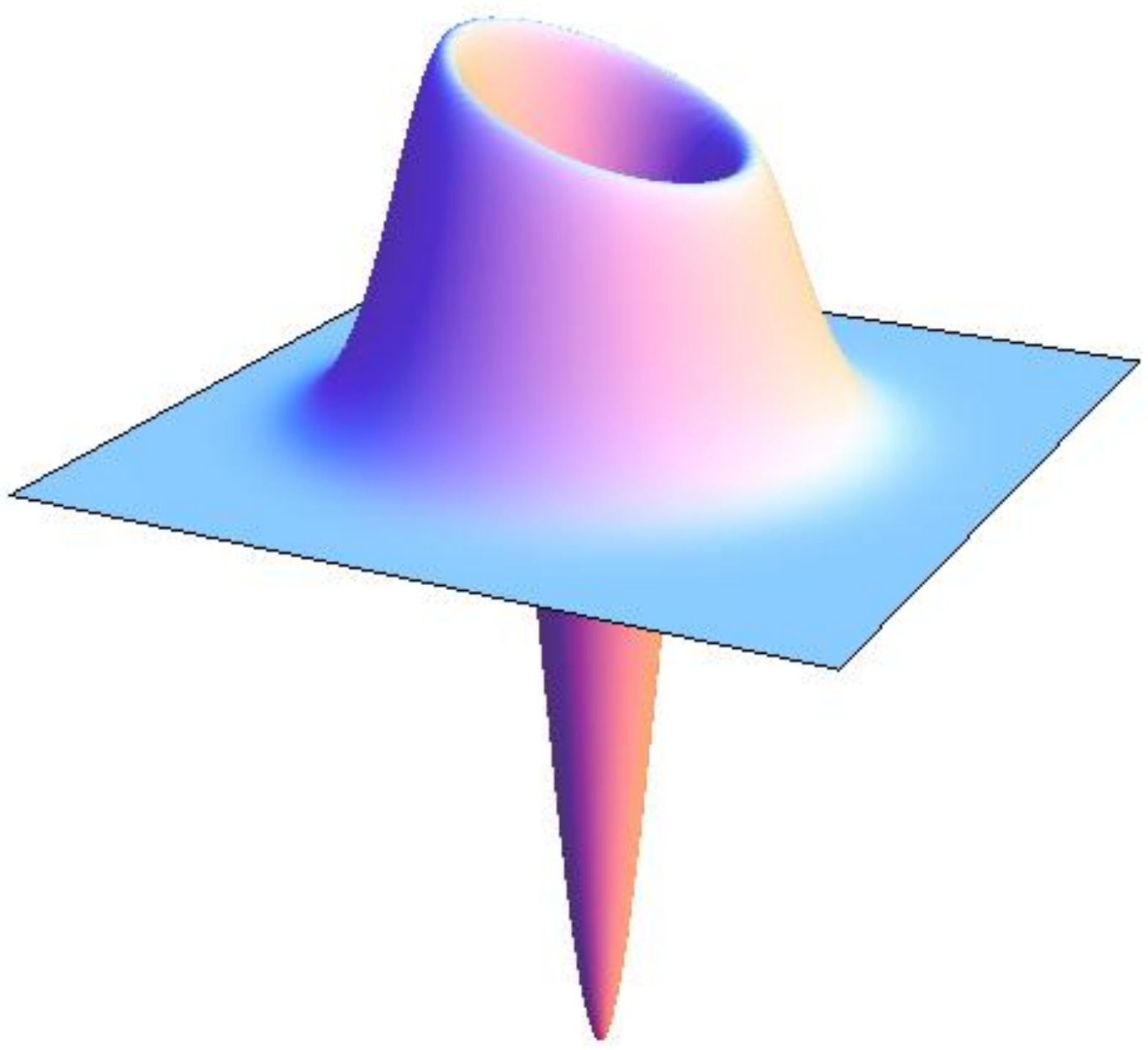} 
\includegraphics[width=0.4\columnwidth]{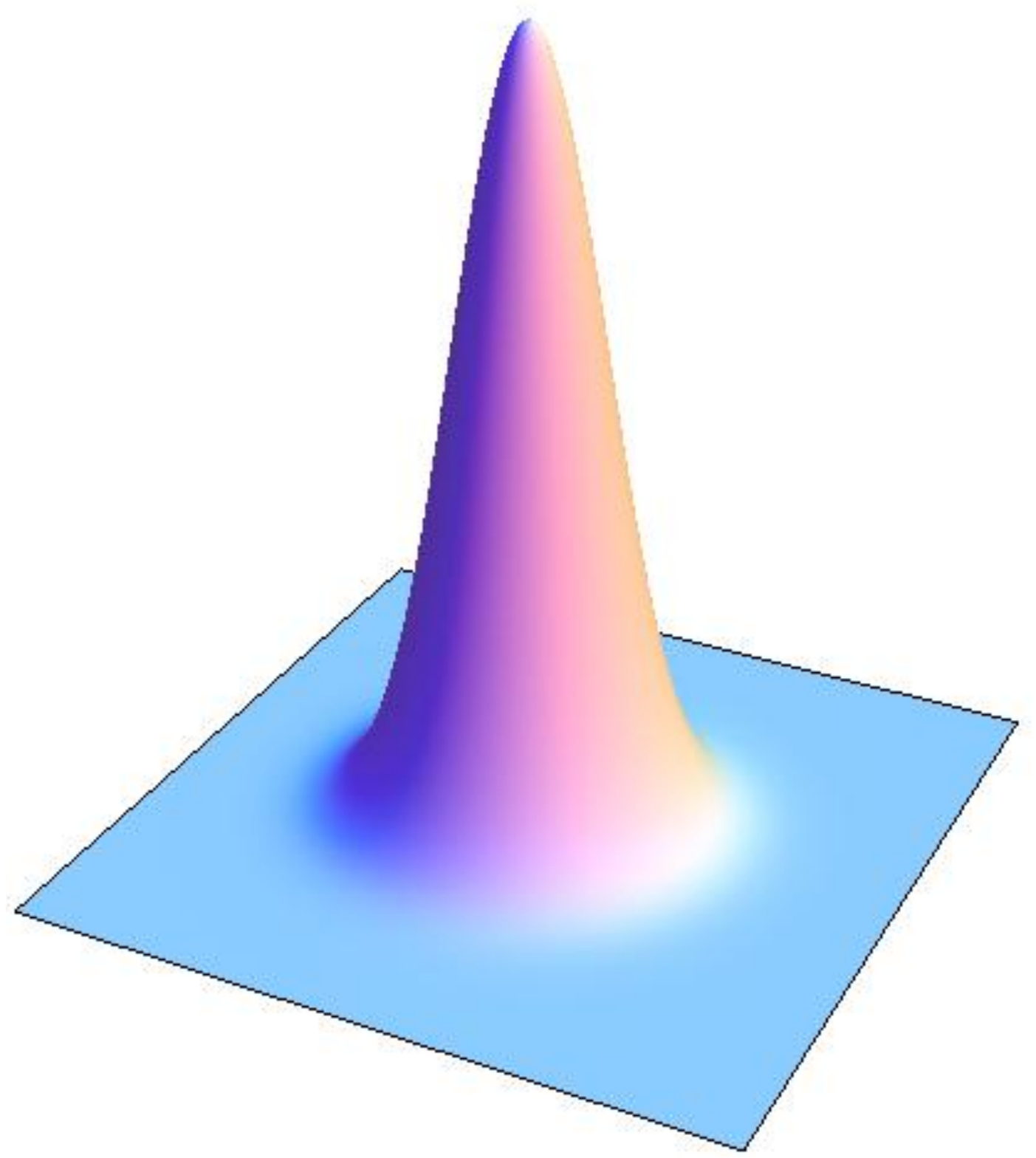}
\caption{(Color online) Wigner function of (left) a SPACS, (right) a coherent state, with amplitude \mbox{$|\alpha|^2=0.01$}. The former is seen to take negative values close to the phase-space origin, while that of the latter is strictly positive everywhere. $W(0)$ is at the center of the plane. Sampling $W(0)$ would distinguish between a coherent state and a SPACS.} \label{wigner}
\end{figure}

\begin{figure}[!htb]
\includegraphics[width=0.75\columnwidth]{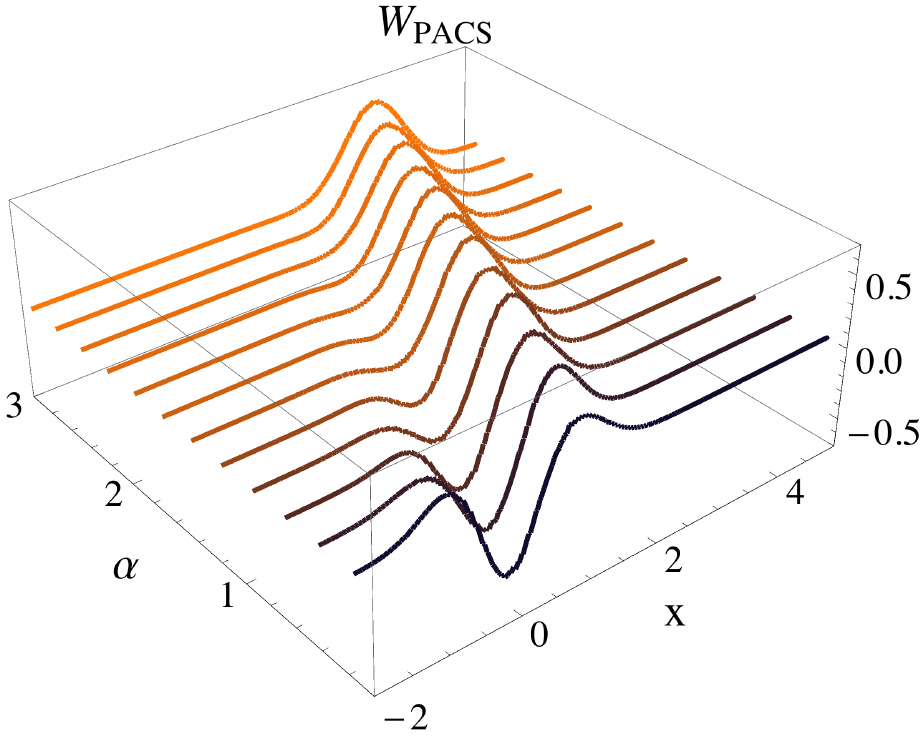} 
\caption{(Color online) 2-d slices of the Wigner function of SPACS across its major axis, as a function of the coherent amplitude $|\alpha|$. We see that the negativity vanishes, and the shape tends towards being a Gaussian for increasing values of $|\alpha|$.}\label{wignerslices}
\end{figure}

The SPACS have been studied extensively in the context of demonstrating quantum-classical transition, since they allow for a seamless interpolation between the highly nonclassical Fock state $|1\rangle$ ($\alpha\rightarrow 0$) and a highly classical coherent state $|\alpha\rangle$ ($|\alpha|>>1$)~\cite{Zavatta_04}. The Wigner function of a SPACS can be expressed as \cite{Agarwal_91}
\begin{equation}
W(z)=\frac{2(|2z-\alpha|^2-1)}{\pi(1+|\alpha|^2)}e^{-2|z-\alpha|^2},
\end{equation}
where $z=x+iy$ is the phase-space complex variable, and $\alpha$ the coherent amplitude in the state. Fig.~\ref{wigner} shows the Wigner functions of a SPACS and a coherent state. The former attains negative values at points close to the origin in phase space, which clearly demonstrates the nonclassical nature of the state. Fig.~\ref{wignerslices} shows a 2-d slice of the Wigner function of a SPACS across the major axis, as a function of the coherent amplitude $|\alpha|$. It can be seen that the Wigner function loses its negativity as $\alpha$ increases and tends towards being a Gaussian state. 

The SPACS-based input that we consider to a linear-optical sampling device can be written as
\begin{align}
\label{input}
|\psi_{\rm in}\rangle^{\mathrm{SPACS}}=&\mathcal{N}\prod_{i=1}^{n}\hat{a}_i^\dagger\hat{D}_i\left(\alpha^{(i)}\right)\ket{0_1,\dots,0_m},\nonumber\\
\mathcal{N}=&\prod_{j=1}^{n}\frac{1}{\sqrt{1+|\alpha^{(j)}|^2}}.
\end{align}
where $\alpha^{(i)}$ represents the complex coherent amplitude in the \mbox{$i$th} mode and $\mathcal{N}$ is the overall normalization factor. That is, the input to the first $n$ modes are SPACS, while the remaining $m-n$ modes are initiated in the vacuum state.  A unitary operation $\hat{U}$ then transforms the state into $|\psi_{\rm out}\rangle^{\mathrm{SPACS}}$
\begin{align}
\label{UxformPACS1}
&=\hat{U}|\psi_{\rm in}\rangle^{\mathrm{SPACS}}\nonumber\\
&=\mathcal{N}\hat{U} \left(\prod_{i=1}^{n}\hat{a}_i^\dagger\hat{D}_i\left(\alpha^{(i)}\right)\right)\hat{U}^{\dagger}\hat{U}\ket{0_1,\dots,0_m}.
\end{align}
This state can be alternatively written as
\begin{align}
\label{UxformPACS2}
%&=\hat{U} \left(\prod_{i=1}^{n}\frac{\hat{a}_i^\dagger\hat{D}_i\left(\alpha^{(i)}\right)}{\sqrt{1+|\alpha^{(i)}|^2}}\right)\hat{U}^{\dagger}\hat{U}\ket{0_1,\dots,0_m},\nonumber\\
&=\mathcal{N}\hat{U} \left\{\prod_{i=1}^{n}\left(\hat{D}_i\left(\alpha^{(i)}\right)\hat{a}_i^\dagger+{\alpha^{(i)}}^*\hat{D}_i\left(\alpha^{(i)}\right)\right)\right\}\hat{U}^{\dagger}\nonumber\\
&\times\ket{0_1,\dots,0_m},
\end{align}
where we have used the commutation relation between the displacement operator and the photon-creation operator, namely
\begin{align}
\left[a^\dagger, \hat{D}(\alpha)\right]=\alpha^*\hat{D}(\alpha).
\end{align} 
We can further simplify the state as
\begin{align}
&=\mathcal{N}\hat{U} \prod_{i'=1}^{n}\hat{D}_{i'}\left(\alpha^{(i')}\right)\hat{U}^{\dagger}\hat{U}\prod_{i=1}^{n}\left(\hat{a}_i^\dagger+{\alpha^{(i)}}^*\right)\hat{U}^{\dagger}\nonumber\\
&\times\ket{0_1,\dots,0_m},\nonumber\\
&=\mathcal{N}\prod_{i'=1}^{n}\left(\hat{U} \hat{D}_{i'}\left(\alpha^{(i')}\right)\hat{U}^\dagger\right)\prod_{i=1}^{n}\left(\hat{U}\hat{a}_i^\dagger\hat{U}^{\dagger}+{\alpha^{(i)}}^*\right)\nonumber\\
&\times\ket{0_1,\dots,0_m}\nonumber\\
&=\mathcal{N}\prod_{j=1}^{m}\hat{D}_{j}\left(\beta^{(j)}\right) \prod_{i=1}^{n}\left(\hat{U}\hat{a}_i^\dagger\hat{U}^{\dagger}+{\alpha^{(i)}}^*\right)\ket{0_1,\dots,0_m},
\end{align}
where $\beta^{(j)} = \sum_{i'} U_{i',j} \alpha^{(i')}$ is the new displacement amplitude in the $j$th mode. Similar to the case of DSPFS sampling, we can now apply a counter-displacement operation of amplitude $\prod_{j=1}^{m}\hat{D}_j\left(-\beta^{(j)}\right)$ (this can be computed efficiently), so that the output state reduces to
\begin{align}
\label{pacsundis_state}
\mathcal{N}\prod_{i=1}^{n}\left(\hat{U}\hat{a}_i^\dagger\hat{U}^{\dagger}+{\alpha^{(i)}}^*\right)\ket{0_1,\dots,0_m}.
\end{align}

Let us denote the state $\prod_{i=1}^{n}\left(\hat{U}\hat{a}_i^\dagger\hat{U}^{\dagger}\right)\ket{0_1,\dots,0_m}$, which corresponds to AA-type boson sampling as $|AA\rangle$. Further, for simplicity, let us choose all the input coherent amplitudes to be equal to $\alpha$. Then, the output state in (\ref{pacsundis_state}) can be written as
\begin{align}
\mathcal{N'}\left(\sum_{i=0}^{n-1}{\alpha^*}^{n-i}\left(\hat{U}\hat{\mathcal{A}}^{(i)}\hat{U}^\dagger\right)\ket{0_1,\dots,0_m}+|AA\rangle\right),
\end{align}
where $\hat{\mathcal{A}}^{(i)}$ is defined for $i\in\{0,1,\cdots,n\}$ as
\begin{equation}
\hat{\mathcal{A}}^{(i)}\equiv
\left\{
	\begin{array}{cl}
		\frac{1}{i!(n-i)!}\sum_{\sigma\in S_n}\prod_{k=1}^{i}\hat{a}^\dagger_{\sigma(k)},  & \mbox{if } i \geq 1 \\
		{\rm id}, & \mbox{if } i =0,
	\end{array}
\right.
\end{equation}
%\begin{align}
%\frac{1}{i!(n-i)!}\sum_{\sigma\in S_n}\prod_{k=1}^{i}\hat{a}^\dagger_{\sigma(k)},
%\end{align}
$S_n$ being the symmetric group of degree $n$, ${\rm id}$ being the identity operator, and $\mathcal{N'}=1/(\sqrt{1+|\alpha|^2})^n$. Now, if we perform photon number detection at the output, the set of all possible outcomes includes total photon numbers (from across all the $m$ output modes) ranging from zero to $n$. Detection events consisting of a total photon number of $n$ would correspond to sampling of the $|AA\rangle$ term from  the superposition. The probability of detecting a total of $i$ photons at the output can be written as
\begin{align}
P_{i}=\mathcal{N'}^{2} {n\choose{i}}\left(|\alpha|^2\right)^{n-i}.
\end{align}
This is because there are ${n\choose{i}}$ terms in $\hat{\mathcal{A}}^{(i)}$, each with a weight of $\mathcal{N'}^{2}\left(|\alpha|^2\right)^{n-i}$.

We now ask the following question: how should $|\alpha|$ scale in terms of $n$---the total number of SPACS in the input (representative of the size of the sampling problem) so that the post-selection probability of detecting $n$ photons at the output of the interferometer scales inverse polynomially in $n$. This is a relevant question to ask, because such a scaling would guarantee the sufficiency of a polynomial number of measurements in order to sample the desired AA term in the output. For simplicity, let us consider ${\rm poly} (n)=n^k$, where $k\in \mathbb{Z}^+$ (the set of positive integers). Solving for $|\alpha|$ that satisfies the above scaling requirement in the limit of a large $n$, we have
\begin{align}
\frac{1}{(1+|\alpha|^2)^n}&\geq \frac{1}{{\rm poly} (n)}\nonumber\\
\Rightarrow 1+|\alpha|^2&\leq ({\rm poly} (n))^{1/n}\nonumber\\
&\leq 1+\epsilon(n),
\label{psp1}
\end{align}
where the third inequality is due to the fact that for all $k\in \mathbb{Z}^+$,
\begin{align}
\lim_{n\rightarrow\infty}(n^k)^{1/n}&=\lim_{n\rightarrow\infty}e^{\frac{k}{n}\log n}\nonumber\\
&=\lim_{n\rightarrow\infty}e^{\frac{k}{n}}=e^{0^+}=1+\epsilon(n).
\label{psp2}
\end{align}
From (\ref{psp1}), we have
\begin{equation}
|\alpha|^2\leq \epsilon(n),
\label{psp3}
\end{equation}
and the large-$n$ expansion
\begin{equation}
e^{\frac{k}{n}\log n}=1+\frac{k}{n}\log n+O(\frac{1}{n^2}),
\end{equation}
tells us that $\epsilon(n)\geq (k/n)\log n$. The chain of inequalities
\begin{align}
\epsilon(n)\geq \frac{k\log n}{n}\geq\frac{1}{n}
\end{align}
thus implies $|\alpha|^2\leq 1/n$ is a sufficient condition on $|\alpha|$ to ensure that the post-selection probability of the AA term scales inverse polynomially in $n$. For $|\alpha|^2=1/n$, in the limit of large $n$, we find that the probability of the term $|AA\rangle$ being detected at the output is
\begin{align}
P_n=\lim_{n\rightarrow\infty}\frac{1}{(1+\frac{1}{n})^n}=\frac{1}{e}\approx 36\%.
\end{align}
%One can therefore efficiently perform post-selection on the number of output photons, so that the measured distribution is entirely consistent with AA boson sampling.  
Further, the probability $P_n$ converges to one when $|\alpha|^2=1/n^2$; i.e., the considered sampling problem with SPACS inputs reduces to AA boson sampling without the need for post-selection. This result is consistent with AA's original result that boson sampling is robust against small amounts of noise. 

On the other hand, we could also ask the question: how should $|\alpha|$ scale, so that the photon number sampling almost always gives the $m$-mode vacuum. For $|\alpha|^2=n^2$, we find that the probability of the $m$-mode vacuum term being detected at the output is
\begin{align}
P_0&=\lim_{n\rightarrow\infty}\frac{{(n^2)}^n}{(1+n^2)^n}\nonumber\\
&=\lim_{n\rightarrow\infty}\frac{1}{(1+\frac{1}{n^2})^n}=1.
\end{align}
That is, the considered sampling problem with SPACS inputs becomes classically simulatable when $|\alpha|^2$ scales as $n^2$, or larger, in the sense that it always results in the detection of the $m$-mode vacuum at the output. 

Therefore, we see that the computational complexity of sampling the SPACS goes from being just as hard as AA's boson sampling for coherent amplitudes $|\alpha|^2\leq 1/n$, to being classically simulatable when $|\alpha|^2\geq n^2$, where $n$ is the total number of SPACS inputs.

\section{Conclusion}
\label{concl}

As discussed in Sec.~\ref{pacssampling}, the SPACS is known to exhibit a quantum-classical transition in terms of the negativity of its Wigner function when the coherent amplitude is tuned from small to large values. The results presented in this work indicate that the sampling problem associated with the SPACS, linear optics and a displaced CPND similarly demonstrates a transition in computational complexity. The complexity goes from being likely hard to simulate classically for small coherent amplitudes (similar to AA boson sampling), to being easy to simulate classically for large coherent amplitudes. This result is also consistent with a conjecture presented in~\cite{GMORD_14} that computational complexity relates to the negativity of the Wigner function.

To summarize, a central open question in the field is what class of quantum states of light yield linear-optics sampling problems that are likely hard to simulate efficiently on a classical computer. Here we have partially elucidated this question by considering two closely related classes of quantum states. We studied the linear-optics sampling of the DSPFS and the SPACS for a displaced CPND. We showed that while DSPFS sampling remains likely hard to simulate efficiently for all values of the displacement, SPACS sampling transitions from being likely hard to simulate efficiently for sufficiently small input coherent amplitudes to being efficiently simulatable in the limit of large coherent amplitudes.

\begin{acknowledgments}
This research was conducted by the Australian Research Council Centre of Excellence for Engineered Quantum Systems (Project number CE110001013). JPD would like to acknowledge the Air Force Office of Scientific Research and the National Science Foundation for support and both KPS and JPD would like to acknowledge the Army Research Office for support. KPS would also like to thank the Graduate School of Louisiana State University for the 2014-2015 Dissertation Year Fellowship.\end{acknowledgments}

\bibliography{paper}

\end{document}